\documentclass[prd,eqsecnum,showpacs,aps]{revtex4}
\usepackage{latexsym}
\begin{document}
\title{Gauge-invariant treatment of the integrated Sachs-Wolfe effect\\
 on general spherically symmetric spacetimes}
\author{Kenji Tomita}
\affiliation{Yukawa Institute for Theoretical Physics, 
Kyoto University, Kyoto 606-8502, Japan}
\date{\today}

\begin{abstract}
On the basis of the Gerlach-Sengupta theory of gauge-invariant
perturbations, a formula of the integrated Sachs-Wolfe effect for a
central observer is derived on general spherically symmetric spacetimes.
It will be useful for comparative studies of theoretical and
observational aspects of the integrated Sachs-Wolfe effect in the
Lemaitre-Tolman-Bondi cosmological models which have been noticed
 by explaining the apparent acceleration without cosmological
constant.
\end{abstract}

\maketitle


\section{Introduction}
\label{sec:level1}
To explain the accelerating behavior of high-redshift supernovas,
isotropic and homogeneous models with nonzero cosmological constant
have been adopted as the standard ones\cite{schm,ries1,ries2,perl},
while we have another possibility to explain the cosmological
acceleration as the apparent phenomenon in spherically symmetric
inhomogeneous models\cite{cele,good,toma,tomb,tomc,tomd}. Recently 
many theoretical studies for their inhomogeneous models have been done
using the Lemaitre-Tolman-Bondi models\cite{igu, yoo, bellido1,
bellido2,bellido3, clif,tomita}.   

While cosmological perturbations in isotropic and homogeneous models
have fully been studied so far\cite{bar,ks}, the treatment of
perturbations in spherically symmetric inhomogeneous models has been
formulated by Gerlach and Sengupta\cite{gs}, and is being studied by
several authors\cite{gund,clark,tomss}. While the integrated
Sachs-Wolfe (ISW) effect also in isotropic and homogeneous models has 
been often studied theoretically and observationally\cite{bou, turok,
tl, tla, coo, is1, is2,ti}, it has been studied in inhomogeneous
models only in a simple case such as being constructed with connected
homogeneous regions\cite{ti2}.   

In this paper we derive a formula of the integrated Sachs-Wolfe effect on
general spherically symmetric inhomogeneous models on the basis of
Gerlach and Sengupta's analysis for the perturbations\cite{gs}. In \S
2 we show the parity classification of perturbations in general spherically
symmetric models and their notations. In \S 3 we consider the 
gauge-invariant perturbations of the null-geodesic equation and the
null condition.  In \S 4 we show a formula of the integrated
Sachs-Wolfe effect in general inhomogeneous models. In \S 5 we
consider the reduction to the isotropic and homogeneous models and
show it can reproduce the well-known formula of the integrated
Sachs-Wolfe effect in the Friedman models.  \S 6 is devoted to the
concluding remarks.

\section{Background model and linear perturbations}
\label{sec:level2}
Following Gerlach and Sengupta's notation\cite{gs}, the metric in
general spherically symmetric spacetimes is expressed as  
\begin{eqnarray}
  \label{eq:a1}
ds^2 &=& g_{\mu \nu} d x^\mu d x^\nu \cr
&=& g_{A,B} d x^A d x^B + r^2 (x^C) d \Omega^2, 
\end{eqnarray} 
where capital Latin indices $A, B, C$ refer to $x^0$ and $x^1$, small
Latin indices $a, b, c$ refer to $x^2$ and $x^3$ (or $\theta$ and
$\phi$), and $d\Omega^2 = d\theta^2 + \sin^2 \theta d \phi^2$.

Linear metric perturbations in the above spacetimes are classified into
odd-parity perturbations and even-parity perturbations. First, the
odd-parity metric perturbations are expressed as @@ 
\begin{equation}
  \label{eq:a2}
h_{\mu \nu} d x^\mu d x^\nu = h_A(x^C) S_a (\theta,\phi)
(dx^Adx^a+dx^adx^A) + h(x^C) (S_{a:b}+S_{b:a}) dx^a dx^b,
\end{equation}
where the covariant derivative of the transverse vector harmonics
$S_a$ on the unit two-surface is indicated by colon $:$. The
gauge-invariant quantities corresponding to $h^A$ and $h$ are 
\begin{equation}
  \label{eq:a3}
k_A = h_A - r^2 (h/r^2)_{,A}.
\end{equation}
Even-parity metric perturbations are
\begin{equation}
  \label{eq:a4}
h_{\mu \nu} d x^\mu d x^\nu = h_{AB}(x^C) Y (\theta,\phi)
dx^Adx^B + h_A(x^C)Y_{,a} (dx^Adx^a+dx^adx^A) +r^2 [K
Y(\theta,\phi)\gamma_{ab} +G Y_{,a:b}] dx^a dx^b,
\end{equation}
and the corresponding gauge-invariant quantities are
\begin{eqnarray}
  \label{eq:a5}
k_{AB} &=& h_{AB} - (p_{A|B} + p_{B|A}), \cr
k &=& K - 2 v^A p_A.
\end{eqnarray}
where $,A$ denotes the partial derivative, $|A$ denotes the
two-dimensional covariant derivative with respect to $x^A$, and
\begin{equation}
  \label{eq:a6}
v_A \equiv r_{,A}/r \quad {\rm and} \quad p_A \equiv h_A -{1 \over 2}
r^2 G_{,A}.
\end{equation}
%

\section{Null-geodesic equations and the null condition}
\label{sec:level3} 
Let us now assume that an observer is at the center of spherical
symmetry and light emitted at the last scattering surface 
reaches the observer in the unperturbed state. Then the light path is
radial and the wave vector $K^\mu (= d x^\mu/d\lambda)$ has the
component $K^A (A= 0, 1)$ and  $K^a (a = 2,3) = 0,$ where $\lambda$ is
the affine parameter. It satisfies the null-geodesic equation
\begin{equation}
  \label{eq:b1}
K^A_{|B} K^B = 0
\end{equation}
and the null condition
\begin{equation}
  \label{eq:b2}
K^A K_A = g_{AB} K^A K^B = 0.
\end{equation}
Their perturbations satisfy the corresponding equations 
\begin{equation}
  \label{eq:b3}
\delta K^\mu_{,\nu} K^\nu + K^\mu_{,\nu} \delta K^\nu + \delta
\Gamma^\mu_{\nu\lambda} K^\mu K^\nu = 0
\end{equation}
and
\begin{equation}
  \label{eq:b4}
\delta g_{AB} K^A K^B + g_{AB} (\delta K^A K^B + K^A \delta K^B) = 0.
\end{equation}
Similarly to the metric perturbations, the perturbation of wave
vector, $\delta K^\mu$ is also classified into the odd-parity and
even-parity perturbations.

The perturbed wave vector of odd-parity $K_{\rm o}$ is
\begin{equation}
  \label{eq:b5}
\delta K_\mu d x^\mu = K_{\rm o} S_a dx^a
\end{equation}
and the wave vector of even-parity have two components $K^{\rm e}$ and
$K^{\rm e}_A$ which satisfy
\begin{equation}
  \label{eq:b6}
\delta K_\mu d x^\mu = K^{\rm e}_A Y dx^A + K^{\rm e} Y_{,a} dx^a.
\end{equation}
For the infinitesimal gauge change of odd parity
\begin{equation}
  \label{eq:b7}
\delta \xi^{\rm odd}_\mu d x^\mu = M(x^C) [-(\sin \theta)^{-1} (\partial
T/\partial \phi) d\theta + \sin \theta (\partial Y/\partial \theta) d
\phi]
\end{equation}
and that of even parity
\begin{equation}
  \label{eq:b8}
\delta \xi^{\rm even}_\mu d x^\mu = \xi_A(x^C) T(\theta, \phi) dx^A +
\xi (x^C) [(\partial Y/\partial \theta) d\theta +(\partial Y/\partial
\phi) d\phi],
\end{equation}
we have
\begin{eqnarray}
  \label{eq:b9}
\bar{K^{\rm o}} - K^{\rm o} &=& - r^2 (M/r^2)_{,A} K^A, \cr
\bar{K^{\rm e}} - K^{\rm e} &=& - r^2 (\xi/r^2)_{,A} K^A , \cr
\bar{K^A_{\rm e}} - K^A_{\rm e} &=& K^A_{,B} \xi^B - K^B \xi^A_{,B}.
\end{eqnarray} 
From these transformation property, we find the gauge-invariant
quantity of odd parity 
\begin{equation}
  \label{eq:b10}
\Psi_{\rm o} \equiv K^{\rm o} + r^2 (h/r^2)_{,A} K^A
\end{equation}
and gauge-invariant quantities of even parity
\begin{eqnarray}
  \label{eq:b11}
\Psi_{\rm e} &\equiv& K^{\rm e} + K^B h_B, \cr
\Phi_{\rm e}^A &\equiv& K^A_{\rm e} - K^A_{|B} p^B + K^B p^A_{|B},
\end{eqnarray} 
where $p^A$ is given in Eq.(\ref{eq:a6}).

Now let us derive the equations for the perturbed wave vectors by
analyzing Eqs. (\ref{eq:b3}) and (\ref{eq:b4}). For odd-parity, we
obtain from Eq. (\ref{eq:b3}) 
\begin{equation}
  \label{eq:b12}
K^{\rm o}_{,B} K^B + 2(r_{,A}/r) K^A K^{\rm o}+ (h_{A,B} -
\Gamma^C_{AB} h_C) K^A K^B = 0.
\end{equation}
By use of gauge-invariant quantities, this equation reduces to
\begin{equation}
  \label{eq:b13}
r^2(\Psi_{\rm o}/r^2)_{,A} K^A = - k_{A|B} K^A K^B,
\end{equation}
where $k_A$ is defined in Eq.(\ref{eq:a3}). For $K^{\rm o}$, we do not
obtain any equation from the null condition Eq.(\ref{eq:b4}).

For even parity, we obtain from the angular part ($\mu = a$) of
Eq. (\ref{eq:b3})  
\begin{equation}
  \label{eq:b14}
K^{\rm e}_{,A} Y^{|a} K^A + \delta \Gamma^a_{AB} K^A K^B 
+ 2 \Gamma^a_{bA} K^A K^{{\rm e}b} = 0,
\end{equation}
where
\begin{eqnarray}
  \label{eq:b15}
\Gamma ^a_{bA} &=& (r_{,A}/r) \delta^a_b, \cr
\delta \Gamma^a_{AB} &=& {1 \over 2}Y^{|a} [-k_{AB} + h_{A|B} + h_{B|A}],
\end{eqnarray} 
so that we obtain
\begin{equation}
  \label{eq:b16}
r^{-2} (\Psi_{\rm e} r^2)_{,A} K^A = {1 \over 2} k_{AB} K^A K^B.
\end{equation}
The integration of this equation along the light path from the emitter
(i) to the observer (f) leads to 
\begin{equation}
  \label{eq:b17}
(\Psi_{\rm e} r^2)_f - (\Psi_{\rm e} r^2)_i = {1 \over 2}
\int^{\lambda_{f}}_{\lambda_{i}} r^2 k_{AB} K^A K^B.
\end{equation}
From the part $\mu = A$ of Eq.(\ref{eq:b3}), we obtain similarly an
equation for $\Phi^A_{\rm e}$ 
\begin{equation}
  \label{eq:b18}
\Phi^A_{{\rm e}|B} K^B + K^A_{|B} \Phi^B_{\rm e} = -{1 \over 2} g^{AB}
(k_{DB|C} + k_{DC|B} - k_{BC|D}) K^B K^C.
\end{equation}
Moreover we obtain from Eq.(\ref{eq:b4})
\begin{equation}
  \label{eq:b19}
\Phi^A_{\rm e} K_A = -{1 \over 2} k_{AB} K^A K^B,
\end{equation}
where $K_A = g_{AB} K^B$. The second term of the left-hand side of
Eq.(\ref{eq:b18}) is rewritten using Eqs.(\ref{eq:b19}) and
(\ref{eq:b1}-\ref{eq:b2}) as
\begin{eqnarray}
  \label{eq:b20}
K^A_{|B} \Phi^B_{\rm e} &=& K^A_{|0} \Phi^0_{\rm e} + K^A_{|1}
\Phi^1_{\rm e}\cr 
&=& (K^A_{|0} - K^A_{|1} K_0/K_1) \Phi^0_{\rm e} -{1 \over 2} k_{BC}
K^B K^C K^A_{|1}/K_1 \cr
&=& -{1 \over 2} k_{BC} K^B K^C K^A_{|1}/K_1,
\end{eqnarray}
so that 
\begin{equation}
  \label{eq:b21}
\Phi^A_{{\rm e} |B} K^B = -{1 \over 2} [g^{AD} (k_{DB|C} + k_{DC|B} -
k_{BC|D}) - k_{BC} K^A_{|1}/K_1] K^B K^C.
\end{equation}
For $A = 0$, the left-hand side of this equation is
rewritten again using Eqs.(\ref{eq:b19}) and
(\ref{eq:b1}-\ref{eq:b2}) as
\begin{eqnarray}
  \label{eq:b22}
(\Phi^0_{{\rm e} ,B} + \Gamma^0_{BC} \Phi^C_{\rm e} ) K^B &=& K^0 
(\Phi^0_{\rm e}/K^0)_{,B} K^B + [K^0_{,B}/K^0 + \Gamma^0_{B0} + \Gamma^0_{B1}
K^1/K^0] \Phi^0_{\rm e} K^B \cr
&& - {1 \over 2} k_{BC} K^B K^C \Gamma^0_{B1} 
K^B/K^1 \cr
&=& K^0 (\Phi^0_{\rm e}/K^0)_{,B} K^B - {1 \over 2} 
K^B K^C \Gamma^0_{B1} K^B/K^1.  
\end{eqnarray}
Therefore we obtain 
\begin{equation}
  \label{eq:b23}
(\Phi^0_{\rm e}/K^0)_{,B} K^B = -{1 \over 2K^0} [g^{0D} (k_{DB|C} +
k_{DC|B} - 
k_{BC|D}) - k_{BC} (K^0_{|1} + \Gamma^0_{D1} K^D)/K_1 ] K^B K^C.
\end{equation}
Since $d(\Phi^A_{\rm e}/K^0)/d\lambda = (\Phi^A_{\rm e}/K^0)_{,B}
K^B$, this equation leads to
\begin{equation}
  \label{eq:b24}
(\Phi^0_{\rm e}/K^0)_f - (\Phi^0_{\rm e}/K^0)_{i} =  -{1 \over 2}
\int^{\lambda_f}_{\lambda_{i}} d\lambda [g^{0D} (k_{DB|C} + k_{DC|B} -
k_{BC|D}) - k_{BC} (K^0_{|1} + \Gamma^0_{D1} K^D)/K_1 ] K^B K^C/K^0,
\end{equation}
and by substituting this solution of $\Phi^0_{\rm e}$ to 
Eq.(\ref{eq:b19}) we can obtain $\Phi^1_{\rm e}$.

\section{Integrated Sachs-Wolfe effect}
\label{sec:level4}
The temperature of the cosmic background radiation measured by an
observer at the center can be written as
\begin{equation}
  \label{eq:c1}
T_o = T_e/(1 + z) = (\omega_o/\omega_e) T_e,
\end{equation}
where $z$ is the redshift of photons during their travel from the
emitter e to the observer o and it is related to the emitted frequency
$\omega_e$ and the observed frequency $\omega_o$. The frequency
$\omega$ measured by the observer with velocity $U^\mu$ is defined
by
\begin{equation}
  \label{eq:c2}
\omega = - g_{\mu\nu} U^\mu K^\nu,
\end{equation}
where $K^\mu = d x^\mu/d \lambda$.

Now we consider the linear perturbations of frequencies, $\delta
\omega$, from the background frequency $\bar{\omega}$. Then $T_o/T_e$
is expressed as
\begin{equation}
  \label{eq:c3}
T_o/T_e = {\bar{\omega}_o + \delta\omega_o \over \bar{\omega}_e +
\delta\omega_e} = {\bar{\omega}_o \over \bar{\omega}_e} ( 1 +
{\delta\omega_o \over \omega_o } - {\delta\omega_e \over \omega_e }).  
\end{equation}
Here we have 
\begin{eqnarray}
  \label{eq:c4}
\delta \omega &=& - \delta g_{\mu\nu} \bar{U}^\mu \bar{K}^\nu -
\bar{g}_{\mu\nu} \delta U^\mu \bar{K}^\nu - \bar{g}_{\mu\nu}
\bar{U}^\mu \delta K^\nu \cr
&=& - \delta g_{AB} \bar{U}^A \bar{K}^B -\bar{g}_{AB} \delta U^A
\bar{K}^B - \bar{g}_{AB} \bar{U}^A \delta K^B,
\end{eqnarray}
because the background quantities have only the components with $\mu =
0$ and $1$. For $\delta g_{AB}$ and $\delta K^A$, we have the
gauge-invariant quantities $k_{AB}$ and $\Phi_{\rm e}^A$. For the
perturbation $\delta U^A$ of the velocity field $U^A$, we can define
the gauge-invariant quantity $V^A$ by
\begin{equation}
  \label{eq:c5}
V^A \equiv U^A_{\rm e} - U^A_{|B} p^B + U^B p^A_{|B}
\end{equation}
with 
\begin{equation}
  \label{eq:c6}
\delta  U_\mu dx^\mu = U_A^{\rm e} Y dx^A + U^e Y_{,a} dx^a,
\end{equation}
where $U^A$ is the unperturbed velocity field, and $U^{\rm e}_A$ and
$U^{\rm e}$ are the components with $\mu = A$ and $a$ of $\delta
U^\mu$, respectively.  
Hence the gauge-independent counterparts of $\delta \omega$ consist
of the following three counterparts
\begin{equation}
  \label{eq:c7}
- k_{AB} \bar{U}^A \bar{K}^B, -\bar{g}_{AB} V^A \bar{K}^B,
-\bar{g}_{AB} \bar{U}^A \Phi^B_{\rm e}.
\end{equation}
The change in frequencies caused by the matter perturbations during
the photon travel from the last scattering surface to the present time 
is given by the last
component which is called the ISW effect. It
should be noticed that only $\Phi^A_{\rm e}$ contributes to the ISW effect
among the three components, while the other components ($\Psi_{\rm o}$ and
$\Psi_{\rm e}$) contribute to the gravitational lensing through
angular deviation of light paths.

Since $\delta \omega/\omega = U_A \delta K^A/U_A K^A$, the
gauge-invariant quantity representing the ISW effect, $\Theta$, is
expressed as
\begin{equation}
  \label{eq:c8}
\Theta \equiv (U_A \Phi^A/U_A K^A)_{\rm f} - (U_A \Phi^A/U_A K^A)_{\rm i}.
\end{equation}

In order to derive $U_A \Phi^A/U_A K^A$ unambiguously in
arbitrary coordinate systems, we
consider the inner products with a system of two orthonormal vectors
$e_A^{(0)}$ and $e_A^{(1)}$, defined as $e_A^{(0)} \equiv U_A$ and
$e_A^{(1)} \equiv n_A$, 
where $n_A$ is the normal vector, i.e. $n_A U^A = 0, U_A U^A
=-1, n_A n^A =1, n_A = g_{AB} n^B$ and $U_A = g_{AB} U^B$. Moreover we
define $e^A_{(B)}$ as $e^B_{(A)} e^{(C)}_B = \delta^C_A$.  
Then we can obtain from Eq.(\ref{eq:b21}) in a similar
manner to the derivation of Eqs.(\ref{eq:b22}) and (\ref{eq:b23})
\begin{equation}
  \label{eq:c9}
(\Phi^{(0)}_{\rm e}/K^{(0})_{,A} K^A = -{1 \over 2K^{(0)}} [g^{(0D)}
(k_{(DB|C)} + k_{(DC|B)} - k_{(BC|D)}) - k_{(BC)} (K^{(0)}_{|(1)} +
\Gamma^{(0)}_{(D1)} K^D)/K_{(1)} ] K^{(B)} K^{(C)},
\end{equation}
where $K^{(A)} = e_B^{(A)} K^B, \ g^{(AB)} = e^{(A)}_C e^{(B)}_D
g^{CD}, \ k_{(AB|C)} = e_{(A)}^D e_{(B)}^E e_{(C)}^F k_{DE|F},$ and
$\Gamma ^{(A)}_{(BC)} = e^{(A)}_D e^E_{(B)} e^F_{(C)} \Gamma^D_{EF}$.
Since we have
$U_A \Phi^A/U_AK^A = \Phi^{(0)}/K^{(0)}$,  $\Phi^{(0)} =
e^{(0)}_A \Phi^A$ and $K^{(0)} = e^{(0)}_A K^A$, we obtain finally    
\begin{equation}
  \label{eq:c10}
\Theta = (\Phi^{(0)}/K^{(0)})_f - (\Phi^{(0)}/K^{(0)})_i =
\int^{\lambda_f}_{\lambda_i} d\lambda [\Phi^{(0)}/K^{(0)}]_{,A} K^A.   
\end{equation}
Using Eqs. (\ref{eq:c9}) and (\ref{eq:c10}), therefore, we can
derive the ISW effect quantitatively.
Here we assume that there are no metric and matter
perturbations at the emission epoch (i) and the observer epoch (f), to
pick up only the integrated component of the Sachs-Wolfe effect
caused by local inhomogeneities between these two epochs.  

\section{Reduction to the homogeneous case}
\label{sec:level5}
As a special case we consider here isotropic and homogeneous
background models
with the metric 
\begin{equation}
  \label{eq:d1}
ds^2 = a^2(\eta) [-d\eta^2 + d\chi^2 +\sigma^2(\chi) d\Omega^2]
\end{equation}
with $\sigma (\chi) = \sin \chi, \chi, \sinh \chi$ for positive, null,
negative curvature. In this case we have $x^0 = \eta$ and $x^1 = \chi$
and the orthonormal vectors are $e^A_{(0)}
= U^A = (1, 0)/a, \ e^A_{(1)} = (0,1)/a, \ e^{(0)}_A = (1,0) a$ and
$e^{(1)}_A = (0, 1) a$. The background wave vectors have the
components $K^0 = - K^1 \propto a^{-2}(\eta)$, or $K^{(0)} = - K^{(1)}
\propto a^{-1}$. In these models the gauge-invariant
scalar-type perturbations are expressed as follows using the
gauge-invariant Newton potential perturbation $\Phi_A$ and spatial
curvature perturbation $\Phi_H$ (in Bardeen's
notation)\cite{bar,ks}. 
\begin{eqnarray}
  \label{eq:d2}
k_{(00)} &=& k_{00}/a^2 = -2 \int d{\bf k} \ \Phi_A Q(\chi), \cr
k_{(01)} &=& k_{01}/a^2 = 0, \cr
k_{(11)} &=& k_{11}/a^2 = 2 \int d{\bf k} \ \Phi_H Q(\chi), \cr
k &=& \int d{\bf k} \ \Phi_H Q(\chi),
\end{eqnarray}
where ${\bf k}$ is the wave number and the scalar harmonics $Q^{(0)}$
are expressed as  
\begin{equation}
  \label{eq:d3}
Q^{(0)} = Q(\chi) Y(\theta, \phi)
\end{equation}
in terms of spherical harmonics $Y(\theta, \phi)$.
Then we obtain from Eqs (\ref{eq:c9}) and (\ref{eq:c10})
\begin{eqnarray}
  \label{eq:d4}
\Theta &=& - {1 \over 2} \int d{\bf k} \int^{\lambda_f}_{\lambda_i} I
K^{(0)} d 
\lambda, \cr 
a I &=& 2 (\Phi_A + \Phi_H)_{,\eta} Q - 4 \Phi_A Q_{,\chi}.
\end{eqnarray}
Here $K^0 d \lambda = d \eta$ or $K^{(0)} d \lambda = a d \eta$. Since
$\chi + \eta = const$ along radial light paths, we obtain
$\int \Phi_A Q_{,\chi} d\eta = \int [-{d \over d\eta} (\Phi_A Q) +
\Phi_{A,\eta} Q] d\eta$, so that 
\begin{equation}
  \label{eq:d5}
\int^{\eta_f}_{\eta_i} a I d\eta = \int^{\eta_f}_{\eta_i} [2 (\Phi_H
-\Phi_A)_{,\eta} Q + 4 {d \over d\eta} (\Phi_A Q) ] d\eta.  
\end{equation}
From Eq. (\ref{eq:d4}) we obtain 
\begin{equation}
  \label{eq:d6}
\Theta = \int d{\bf k} \int ^{\eta_f}_{\eta_i} (\Phi_A
-\Phi_H)_{,\eta} Q d \eta. 
\end{equation}
It should be noticed here that another term $2 \int d{\bf k} \ (\Phi_A
Q)|^f_i$ in Eqs. (\ref{eq:d4}) and (\ref{eq:d5}) vanishes, because the
influence from metric 
perturbations at the emission and observer epochs are neglected.
The above result is well-known as the ISW 
effect in the Friedman models\cite{gou}.

\section{Concluding remarks}
\label{sec:level6}
We derived a formula of the ISW effect for a central observer which
is appropriate to studying the ISW effect caused by local
inhomogeneities like galaxies, clusters and
voids between the emission and observer times. The
next step is to extend this formula to the case of off-central
observers, which will be more complicated because of the additional
asymmetry. 

When the metric perturbations in realistic Lemaitre-Tolman-Bondi
 models will be
explicitly obtained, our formula will be useful to derive the ISW
effect in them theoretically and compare it 
to the observational one. As another next step, we shall attempt to derive
it in a case of self-similar spacetimes.  




\end{document}